%% file: coriolis2.tex
\documentstyle[eqsecnum,aps,bezier,amsmath,epsfig]{revtex}

\begin{document}

\twocolumn[\hsize\textwidth\columnwidth\hsize\csname@twocolumnfalse\endcsname

\title{Coriolis force in Geophysics: an elementary introduction and examples}
\author{F. Vandenbrouck$^\star$, 
L. Berthier$^{\star\star}$, and
F. Gheusi$^{\star\star\star}$}
\address{$^\star$Laboratoire de Physique de la Mati\`ere Condens\'ee, 
Coll\`ege
de France, 11 place M. Berthelot, 75005 Paris, France}
\address{$^{\star\star}$Laboratoire de Physique, ENS-Lyon and CNRS, 
46 all\'ee d'Italie, 69007 Lyon, France}
\address{$^{\star\star\star}$M\'et\'eo-France, CNRM/GMME/Relief, 42 avenue 
G. Coriolis,
31057 TOULOUSE Cedex, France}

\date{\today}

\maketitle
\begin{abstract}
We show how Geophysics may illustrate and thus improve 
classical Mechanics lectures concerning the study of
Coriolis force effects.
We are then interested in atmospheric as well as oceanic phenomena we are
familiar with, and are for that reason of pedagogical and practical interest.
Our aim is to model them in a very simple way to bring out the 
physical phenomena that are involved.

{\bf LPENSL-TH-06/2000}.
\end{abstract}

\twocolumn\vskip.5pc]\narrowtext

\section{Introduction}

The study of mechanics in non-inertial frames, the 
Coriolis force effects for the rotating Earth
being the paradigm, are 
often restricted to the standard examples of a deflected projectile
and the Foucault pendulum. 
In these two examples, the Coriolis force is only a {\it small perturbation}.
In order to make the Coriolis force effects dominant,
one must consider phenomena taking place
at the geophysical scale.

This is the line followed by the present paper.
The first section is devoted
to the presentation of the basic equations. 
In the second section, we discuss the Physics
of anticyclones and cyclones: we describe their rotation in the
geostrophic approximation, and show how cyclones (but not
anticyclones) may become hurricanes. 
The third section presents
a second  example of atmospheric circulation, the Jets Streams, which
are stabilized by the Coriolis force. 
We show also that these strong
winds are associated with Rossby waves. 
The last section presents two
examples of oceanic circulation: wind-driven circulation and
Kelvin waves.

\section{Basic equations}

\subsection{Navier-Stokes equation in a rotating frame}

Let us introduce two frames ${\cal R}$ and ${\cal R}^{\prime}$ in
relative motion. The inertial frame ${\cal R}$ is the geocentric
one, with origin at the center $O$ of the Earth, and whose axes are
along directions of fixed stars. The frame ${\cal R}^{\prime}$ is
fastened to the Earth. 
It has an angular velocity
$\boldsymbol{\Omega}$ with respect to ${\cal R}$, where
$\boldsymbol{\Omega}$ is the angular velocity of rotation of the
Earth ($\Omega \simeq 7.29\cdot10^{-5}\,{\mathrm {rad}}\cdot{\mathrm
s}^{-1}$). The following relation between the accelerations
of a point $M$
$\boldsymbol{a}_{\cal R}(M)$ in ${\cal R}$, and $\boldsymbol{a}_{{\cal
R}^{\prime}}(M)$ in ${\cal R}^{\prime}$ may easily be
obtained~\cite{goldstein}:
\begin{equation}
\boldsymbol{a}_{\cal R} (M) = \boldsymbol{a}_{{\cal R}^{\prime}}(M)
 + 2\boldsymbol{\Omega} \wedge
\boldsymbol {v}_{{\cal R}^{\prime}}(M) + \boldsymbol{\Omega} \wedge
 (\boldsymbol{\Omega}
\wedge\boldsymbol{OM}).
\label{cinemat}
\end{equation}
In eq.~(\ref{cinemat}), the term $2\boldsymbol{\Omega} \wedge
\boldsymbol{v}_{{\cal R}^{\prime}}(M)$ is the Coriolis acceleration,
$\boldsymbol{v}_{{\cal R}^{\prime}}(M)$ is the velocity of 
$M$ in ${\cal R}^{\prime}$,
and $\boldsymbol{\Omega} \wedge (\boldsymbol{\Omega}
\wedge \boldsymbol{OM})$ is the centrifugal acceleration. In the
rotating frame ${\cal R}^{\prime}$, the Navier-Stokes equation
takes into account the above inertial terms and reads~\cite{green}:
\begin{equation}
\begin{aligned}
\frac{\partial \boldsymbol{v}_{{\cal R}^{\prime}}}{\partial t} +
(\boldsymbol{v}_{{\cal R}^{\prime}}
\cdot \boldsymbol{\nabla})\boldsymbol{v}_{{\cal R}^{\prime}} = &
-\frac{1}{\rho}\boldsymbol{\nabla}p+\frac{1}{\rho}\boldsymbol{f} -
2\boldsymbol{\Omega}\wedge\boldsymbol{v}_{{\cal R}^{\prime}} \\ & -
\boldsymbol{\Omega}\wedge(\boldsymbol{\Omega}\wedge\boldsymbol{OM})
+\nu\Delta\boldsymbol{v}_{{\cal
R}^{\prime}}.
\label{ns}
\end{aligned}
\end{equation}
The force $\boldsymbol{f}$ includes the gravitational force and
other external forces if they exist, $\rho$ is the
density of the fluid and $p$ the pressure field.
The dependence on $M$ has been removed in all the terms for clarity.
 The centrifugal force is conservative.
If this is also the case for $\boldsymbol{f}$, one can rewrite the
terms $\boldsymbol{\nabla}p$, $-\boldsymbol{f}$ and $\rho
\boldsymbol{\Omega}  \wedge(\boldsymbol{\Omega} \wedge
\boldsymbol{OM})$ as $\boldsymbol{\nabla}p^{\prime}$, where
$p^{\prime}$ is called {\it dynamical pressure}. In the rotating
frame, the hydrostatic equilibrium equation is:
$\boldsymbol{\nabla}p^{\prime}\,=\,\boldsymbol{0}$. The dynamical
pressure $p^{\prime}$ reads, within a constant, $p^{\prime} \,=\,p
+
\rho g z$, where $g$ is the Earth gravity field.
Recall that $g$ includes the
centrifugal term, and is thus slightly different from
the gravitational field, which only takes into 
account the Earth's attraction~\cite{goldstein}.

\subsection{Reynolds and Rossby numbers}

The nonlinearity of the Navier-Stokes equation makes it
difficult to solve in general. 
It is hence necessary to evaluate
the relative importance of the different terms in order to
make further simplifications. 
This is done by introducing 
the different characteristic scales of the
flow: $L$ denotes the typical spatial extension, $U$
the velocity, $\Omega$ the angular velocity and $\nu$ the kinematic
viscosity. Two non-dimensional numbers may then be derived
from these scales.

(i) The Reynolds number is defined as:
\label{reynono}
\begin{equation}
R_e  = \left| \frac{(\boldsymbol{v} \cdot \boldsymbol{\nabla})
 \boldsymbol{v}}
{\nu \Delta \boldsymbol{v}} \right| = \frac{U^2 / L}{\nu U / L^2} =
\frac{UL}{\nu}.
\end{equation}
It characterizes the relative importance of the
momentum transport in the fluid through advection and
viscous diffusion. For the atmospheric flows studied
here, typical values are: $ U\sim 10\ {\mathrm m}\cdot{\mathrm
s}^{-1}$, $L \sim 10\ {\mathrm {km}}$ and $\nu\sim10^{-5}\ {\mathrm
m}^2\cdot{\mathrm s}^{-1}$. Thus, the Reynolds number is about
$10^{10}$. A large value of the Reynolds number is also obtained
for oceanic flows~\cite{note}. Hence, the
 Navier-Stokes equation reduces, for
geophysical flows, to the Euler equation:
\begin{equation}
\begin{aligned}
\frac{\partial \boldsymbol{v}_{{\cal R}^{\prime}}}{\partial t} +
(\boldsymbol{v}_{{\cal R}^{\prime}}
\cdot \boldsymbol{\nabla})\boldsymbol{v}_{{\cal R}^{\prime}} = &
-\frac{1} {\rho}\boldsymbol{\nabla}p 
  + \frac{1} {\rho}\boldsymbol{f}\\ & -2 \boldsymbol{\Omega} \wedge
\boldsymbol{v}_{{\cal R}^{\prime}} - \boldsymbol{\Omega}
\wedge(\boldsymbol{\Omega}\wedge\boldsymbol{OM}).
\label{euler}
\end{aligned}
\end{equation}
Moreover, geophysical flows are turbulent (high Reynolds 
number)~\cite{lesieur}.
For the sake of simplicity, we ignore this complication in what
follows. A simple way of taking into account the relevant effects
of turbulence will be presented in the last section. (See section
\ref{sec:ekman}.)

(ii) The Rossby number is defined as:
\begin{equation}
R_o \,=\,\left|\frac
{(\boldsymbol{v}\cdot\boldsymbol{\nabla})\boldsymbol{v}}
{2\boldsymbol{\Omega}\wedge\boldsymbol{v}}\right|\,=\,\frac{U^2 /
L}{\Omega U}\,=\,\frac{U}{L \Omega}.
\label{Rossby}
\end{equation}
It compares the advection and
the rotation effects. 
The Coriolis force dominates 
if $R_o\ll1$. A geophysical flow, characterized
by a large spatial extension, may easily be influenced by the
Earth's rotation, as one typically has $R_o \sim10^{-2}
\, \ll  \,1$. 
On the other hand, an emptying bathtub with 
$U \sim 1\,{\mathrm m}\cdot{\mathrm s}^{-1}$, and $L \sim
10^{-1}{\mathrm m}$, has $R_o \sim 10^5$.
Such a
flow is  more strongly influenced by the advection in the fluid, and
thus by the initial conditions, than by the Earth's
rotation.

\section{Atmospheric eddies}

\subsection{Anticyclones and cyclones}
\label{geoo}

We consider first the situation when the Rossby number is negligible.
This is the case for anticyclones and cyclones since one typically
has $ U\sim 10\ {\mathrm m}\cdot{\mathrm s}^{-1}$, $L \sim 1000
\, \mathrm{km}$, which yields $ R_o \sim 0.1$. In the Euler
equation~(\ref{euler}), we only have to keep the gravity, pressure
and Coriolis terms. This hypothesis constitutes the {\it
geostrophic approximation}. For each point $M$ of the Earth, we
define a vertical axis ($Mz$), and a cylindrical coordinate system
$(r,\theta,z)$. The vertical component of the velocity field is
supposed to be zero, which implies that the movements of the fluid are
locally horizontal. $u$ is the radial component of the
velocity field and $v$ the tangential one. The Earth's angular
velocity $\boldsymbol{\Omega}$ is written as $\boldsymbol{\Omega}
\,= \,\boldsymbol{\Omega_\parallel} +\boldsymbol{\Omega_\perp}$
where $\boldsymbol{\Omega_\parallel} \, \equiv \,  \Omega \sin
\lambda \, \boldsymbol{u_z}$ and $\boldsymbol{\Omega_\perp}$ is
$\boldsymbol{\Omega}$'s projection on the plane ($r,\theta$);
$\lambda$ is the latitude. The flow is supposed to be stationary.
In this system of coordinates, the Euler equation can be rewritten,
under the geostrophic approximation, as:
\begin{mathletters}
 \label{geo}
 \begin{eqnarray}
        \frac{\partial p}{\partial r} & = & \rho v f
        \label{geoa},\\
        \frac{1}{r}\frac{\partial p}{\partial \theta} & = &-\rho u f
        \label{geob},\\
       \frac{\partial p}{\partial z} & = & -\rho g - 2\rho
       (\boldsymbol{\Omega_\perp}\wedge\boldsymbol{v})\cdot\boldsymbol{u_z}.
         \label{geoc}
 \end{eqnarray}
\end{mathletters}
In these equations,
$f \equiv 2 \Omega \sin \lambda$ is the {\em Coriolis parameter}.
 In equation (\ref{geoc}), the term $2\rho (\boldsymbol{\Omega_\perp}
\wedge \boldsymbol{v}) \cdot \boldsymbol{u_z}$ is small compared to
$\rho g$ ($\Omega U / g \sim 10^{-5}$). Equation (\ref{geoc})
therefore reduces to the hydrostatic equilibrium equation
$\partial p /
\partial z \, = \, - \rho g$.

If we consider the case of an eddy in the Northern
hemisphere and assume that the velocity field is tangential 
($u = 0$), then, $v<0$ (clockwise rotation) implies $\partial p/\partial
r<0$.
The pressure is higher at the eddy center than outside: it is an anticyclone. 
A cyclone would correspond to an
anticlockwise rotation. Both situations are represented in
figure \ref{fig:anti}. The rotation senses are opposite
in the southern hemisphere.

\begin{figure}
\psfig{file=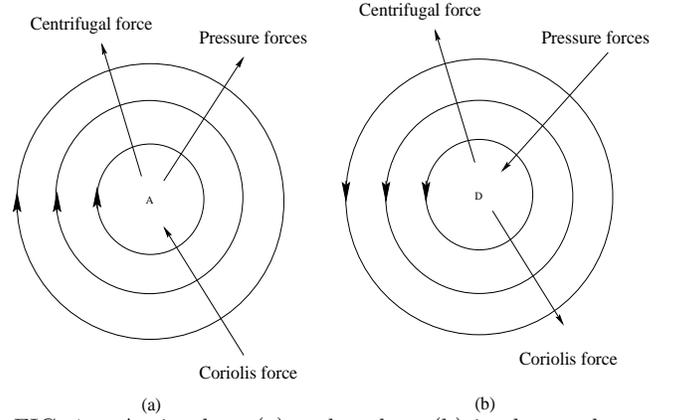,width=8.5cm,height=5.5cm}
\caption{\label{fig:anti} Anticyclone (a) and cyclone (b) in the northern
hemisphere.}
\end{figure}

We end this section with two concluding remarks about the presence
of the Coriolis force:

(i) Without this force, an eddy center is {\it always} a pressure minimum.
However, in the case of the anticyclone, the Coriolis force stabilizes 
the inverse situation, with the eddy center being a pressure maximum.

(ii) In its vectorial form, the geostrophic
 equilibrium equation reads: $\boldsymbol{\nabla}
p^{\prime} \, = \, -2 \rho \boldsymbol{\Omega} \wedge \boldsymbol{v}$.
This implies that the pressure $p^{\prime}$ is constant along a 
streamline. When the usual Bernoulli equation is valid, pressure 
variations are, on the contrary, associated with velocity 
variations along a streamline. 

\subsection{Hurricanes}

Let us consider an eddy (anticyclone or cyclone) whose
 angular velocity and radius are 
respectively $\omega$ and $R$. The Rossby
number characterizing this eddy can be written as $R_o \,=
\, U / L\Omega \,= \, \omega /\Omega$. Therefore, the geostrophic
equilibrium corresponds to a small angular velocity of
the eddy, {\it i.e.} $\omega \ll \Omega$. We shall now consider the
case where the eddy's angular velocity is not small compared to the
Earth's rotation. This means that the centrifugal force due to the
eddy rotation has to be taken into account. In this case, the
Rossby number is of order unity. In the frame ${\cal R}^\prime$, the
fluid particle has a uniform circular motion. Forces acting on it
are the Coriolis force and the radial pressure gradient. The
equation of motion for a fluid particle, located at the eddy's
periphery reads, in ${\cal R}^\prime$:
\begin{equation}
       -r_0\omega^2\,=\,-\frac{1}{\rho}\frac{{\mathrm d}p}{{\mathrm
       d}r}+r_0f\omega,
       \label{equicyclone}
\end{equation}
where $r_0$ is the eddy radius. The term $-r_0\omega^2$ corresponds
to the centrifugal acceleration of the fluid particle, and
$r_0f\omega$ is the Coriolis term.

An anticyclone in the northern hemisphere is shown in 
figure \ref{fig:anti}a. For such an equilibrium, the Coriolis
 force compensates 
both pressure and
centrifugal forces. If the angular velocity of the anticyclone
grows, the Coriolis force is not sufficient to counterbalance
these two forces since the centrifugal force grows faster than the
 Coriolis force 
with increasing $\omega$. This is not the case for the cyclone depicted
in the figure \ref{fig:anti}b. The pressure and centrifugal
forces may counterbalance each other when the rotation of the
cyclone becomes faster. This qualitative approach shows that
there is no limit to the kinetic energy of rotation for a
cyclone.

More quantitatively, equation~(\ref{equicyclone}) can be solved to
find:
\begin{equation}
    \omega_{\pm} = \frac{f}{2} \left[ -1 \pm \sqrt{ 1 + \frac{G}{G_0}} \right],
    \label{racines}
\end{equation}
where $G\equiv dp/dr$ and $G_0 \equiv \rho r_0
f^2 / 4$. Figure \ref{fig:omega} gives the
evolution of an eddy angular velocity as a function of the radial
pressure gradient. In this figure, the geostrophic situation can be
found around the origin (small pressure gradient and angular
velocity). In the northern hemisphere, the sign of the angular
velocity is the same as that of the pressure gradient. 
One can even obtain
the angular velocity of an eddy by developing the
expression~(\ref{racines}) around zero: $\omega \approx fG/4G_0$.
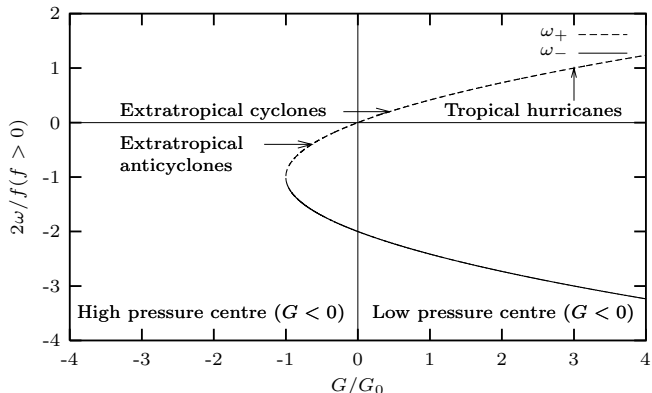
\begin{figure}
\input{omega2}
\caption{\label{fig:omega} Normalized angular velocity as a
 function of the normalized pressure
gradient.}
\end{figure}
The condition $G>-G_0$, for the existence of the above solutions,
gives a limit to the angular velocity of an anticyclone
($G<0$). One finds $\omega_{max} = 2\Omega\sin\lambda$.
This limit does not exist for a cyclone
($G>0$). When the angular velocity grows, the radial pressure gradient
follows this evolution and becomes more and more
important. This explains why hurricanes are always associated with
very low pressure. 

We note in conclusion that the balance between 
the centrifugal force and the 
radial pressure gradient is possible 
whatever the sense of rotation. 
Thus, the existence of clockwise 
hurricanes in the northern hemisphere cannot be excluded. 
However, most of the hurricanes observed in the northern hemisphere are 
anticlockwise and result from the amplification of earlier 
tropical cyclones, 
the amplification mechanism 
being the conversion of the latent heat 
of evaporating tropical warm waters into rotational kinetic energy.

\section{Jet Streams and Rossby waves}

The difference in solar heating between the equatorial
 and polar regions drives 
a convective cell at the planetary length scale, 
the {\em Hadley cell}, which extends 
in both hemispheres from the Equator up to the sub-tropical latitudes.
 The heated 
equatorial air rises, moves toward the poles 
where it cools, then sinks and comes back to 
the equator.
When coming back, the air is 
deflected toward the west 
by the Coriolis force, generating easterlies at the tropical
latitudes which are known as 
the {\em Trade Winds}. 
Conversely, the upper-troposphere
trajectories toward the poles
are deflected toward the east. 
Because of the thermal structure of the 
atmosphere~\cite{holton}, these 
upper-level westerlies concentrate in narrow tubes of very strong
 winds up to 
$80\ \mathrm{m}\cdot\mathrm{s}^{-1}$, the {\em Jet Streams}.
 The Jet Streams are typically found at 
altitudes of about $10\ \mathrm{km}$ and at latitudes between $30^\circ$
 and $40^\circ$. However, 
their strength and location may depart significantly from these mean values, 
depending on the season, the longitude, and the day-to-day thermal
 structure of 
the atmosphere at mid latitudes. It can be noted that B. Piccard 
and B. Jones took great advantage of the Jet Streams for their 
recent travel around the world in a balloon. The Jet Streams are
also useful to the 
planes flying from America to Europe. 

In this section, we propose to show how a {\it zonal} wind ({\it
i.e.} along the parallels) may be stabilized by the Coriolis force.
A mass $M$ of air near the Earth's surface is reduced to a point
$G$. Its coordinates are the usual spherical ones
($R,\theta,\varphi$), $\theta$ being the colatitude and $R$ the
radius of the Earth. The velocity of $G$ can then be explicitly
written: $\boldsymbol{v}_{{\cal R}^\prime}(G) \, = \,
R\dot{\theta}\,\boldsymbol{u_\theta}+R\dot{\varphi}
\sin\theta\,\boldsymbol{u_\varphi}$.
The quantity $R\dot{\varphi}\sin\theta$ is the drift velocity $u_0$
of the point $G$ along a parallel. We deduce the following expression
of the Coriolis force moment about the centre of the Earth (point
$O$):
\begin{equation}
       \boldsymbol{{\cal
       M}}_O \, =\,  2MR^2\Omega\dot{\theta}\,\boldsymbol{u_\theta}+2MR\Omega
u_0\cos\theta\,\boldsymbol{u_\varphi}.
       \label{moment}
\end{equation}
The computation of the angular momentum of $G$ about $O$, in the frame
${\cal R}^ \prime$, yields:
\begin{equation}
       \boldsymbol{L_{{\cal
       R}^\prime}}(O)  =
-(MR^2)\dot{\varphi}\sin\theta\,\boldsymbol{u_\theta}
    +(MR^2)\dot{\theta}\,\boldsymbol{u_\varphi}.
       \label{momcin}
\end{equation}
The theorem of angular momentum for the point $G$, about $O$ and
projected  on $\boldsymbol{u_\varphi}$ gives:
\begin{equation}
       -\ddot{\lambda} \,=\, 2 \frac{\Omega u_0}{R} \sin\lambda,
\end{equation}
where $\lambda \equiv \pi/2-\theta$ is the latitude.
This equation is linearized for small deviations around a given latitude 
$\lambda_0$, leading to
\begin{equation}
       \ddot{\delta\!\lambda} + \left[ 2\frac{\Omega
       u_0}{R}\cos\lambda_0 \right] \,\delta\!\lambda = 0,
\end{equation}
where $\delta\!\lambda \equiv \lambda-\lambda_0$.
The meridional motion of $G$ remains bounded, only if $u_0>0$, 
which corresponds to a drift velocity from west to east.
This motion is characterized by small oscillations around
the mean latitude $\lambda_{0}$ with angular frequency
$\omega_0\,=\,\sqrt{2\Omega u_0\cos\lambda_0/R}$. These
oscillations correspond to the {\it stationary case of a Rossby
wave}~\cite{rossby}. 
More generally, Rossby waves in the atmosphere are guided by strong westerlies.

\section{Oceanic circulation}

Oceanic circulation is, of course, described by the same equations as
atmospheric circulation. 
For large scale oceanic currents, like {\it
e.g.} the Gulf stream, the geostrophic approximation (see
section~\ref{geoo}) is relevant: the Coriolis force compensates
the horizontal pressure gradient, which is related to the slope 
of the free surface, 
which is not necessarily horizontal~\cite{openuniv1}.

We shall be interested here in a slightly different case for which
the interaction between the wind and the ocean gives rise to a
current.

\subsection{Wind-driven circulation : Ekman transport}
\label{sec:ekman}

The wind induces a friction at the ocean surface,
transmitted through turbulence to the deeper layers of the sea.
There is a supplementary difficulty that we cannot ignore here. 
The flow is not laminar, but
essentially turbulent.
The fluid viscosity is related to molecular agitation, dissipating
the energy of a fluid particle. A diffusive momentum transport is
associated with this phenomenon. In a turbulent flow, agitation
dissipates the energy associated with the mean velocity of the
current. This analogy allowed Prandtl to introduce the notion of an
{\it eddy viscosity}~\cite{lesieur}. 
In this approximation, considering 
$\boldsymbol{v}_{{\cal R}^{\prime}}$ as the mean
flow velocity, the Navier-Stokes
equation (\ref{ns}) remains unchanged, the eddy viscosity
$\nu_{turb}$ being added to the kinematic viscosity $\nu$. 
It must be remarked that the
former is a property of the flow while the latter is a property of the fluid. 
As far as geophysical flows are concerned, the kinematic
viscosity is neglected, since typically $\nu_{turb}/\nu \sim
10^{5}$ for oceanic flows, and $\nu_{turb}/\nu$ is about $10^{7}$
for atmospheric flows.

Let us write the Navier-Stokes equation in projection on $(Oxyz)$,
where $(Oxy)$ is the surface of the globe, $(Oz)$ the ascendant
vertical, and $(u, v, w)$ are the velocity components:
\begin{equation}
\begin{aligned}
        \frac{du}{dt} & =   - \frac{1}{\rho}
 \frac{\partial p}{\partial x} + fv + \nu_{turb} \left(
\frac{{\partial}^{2} u}{\partial x^{2}} +
\frac{{\partial}^{2} u}{\partial y^{2}} + \frac{{\partial}^{2}
u}{\partial z^{2}} \right),\\
        \frac{dv}{dt} & =  - \frac{1}{\rho}\frac{\partial p}{\partial y} - fu +
\nu_{turb} \left(  \frac{{\partial}^{2} v}{\partial x^{2}} +
\frac{{\partial}^{2} v}{\partial y^{2}}+ \frac{{\partial}^{2}
v}{\partial z^{2}} \right).
         \label{projection}
\end{aligned}
\end{equation}
For a stationary situation with large Rossby number,
the acceleration terms are negligible: the velocity depends then 
only on space. 
The horizontal pressure
gradient terms can also be neglected since the equations have
been linearized and 
one can consider the real
physical situation as the superposition of a geostrophic current
(taking into account the pressure terms) and a wind-driven current,
which will now be described. We consider a solution depending on
space only through the coordinate $z$.
The boundary conditions are the following: 
the velocity has to be finite both as $z \rightarrow 
-\infty$ and at the free surface, the stress is 
proportional to $\partial \boldsymbol{v} / \partial
z$ 
and parallel to the wind flow, assumed to be in the ($Oy$) direction.
One can solve eq.(\ref{projection}) and find the velocity field 
(the solution is
straightforward defining $W(z) \equiv u(z) +iv(z)$):
\begin{equation}
\begin{aligned}
u(z) & =  \pm V_{0} \cos \left( \frac{\pi}{4} + \frac{z}{\delta}
\right) \exp \left( \frac{z}{\delta} \right), \\
 v(z) & =  V_{0} \sin \left( \frac{\pi}{4} +
 \frac{z}{\delta} \right) \exp \left(\frac{z} {\delta}
\right),
\end{aligned}
\end{equation}
where $\delta \equiv \sqrt{2 \nu_{turb} / |f|}$ is a distance
called the {\it Ekman depth}. Typical values for $\delta$ are $\delta
\sim 10\,-\,100 \, \mathrm{m}$. ``$+$'' stands for the northern
hemisphere and ``$-$'' for the southern one.
\begin{figure}
\psfig{file=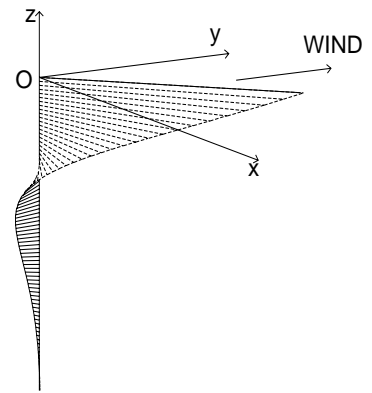,width=8.5cm,height=6.5cm}
\caption{\label{spirale}Ekman spiral.
 The surface is generated by the velocity field
$\boldsymbol{v}(0,0,z)$.}
\end{figure}
Close to the surface ($z=0$), the current deviates $45^{\circ}$,
 and the direction of the velocity
rotates clockwise
(anticlockwise) in the northern (southern) hemisphere. The amplitude of 
the velocity decreases exponentially on a
length scale $\delta$, which represents the characteristic depth 
over which
the influence of the wind is significant. This velocity field,
 the so-called {\it
Ekman spiral}, is plotted in figure~\ref{spirale}.
The mean effect of the wind, over a depth $\delta$, 
is the fluid motion in a direction perpendicular
to it:
this effect is called the {\it Ekman transport}.

\subsection{Kelvin waves}

The main difference between atmospheric flows and oceanic flows
occurs near in the coastline, limiting the waters motion.
This is the origin of {\it Kelvin waves}.
If one considers the deformation of the free surface of the oceans, one can 
see that gravity acts as a restoring force, giving rise to a 
``gravity wave''~\cite{landau}. When influenced by the Earth's 
rotation, these waves are called ``modified waves''~\cite{openuniv2}.

Let us consider the following geometry: a south-north current, with
a coast on its right (east). The coast is supposed to be a vertical
wall, the water height being denoted $h_{0}+h(x,y,t)$. The Coriolis
force usually deflects a south-north current toward the 
east, {\it i.e.} toward the coast. 
Hence water gathers close to the coast, and gives rise to an west-east 
horizontal pressure gradient counterbalancing the Coriolis force.
The equations describing the gravity waves are
 the linearized Euler and continuity
equations~\cite{landau}:
\begin{equation}
\begin{aligned}
\frac{\partial u}{\partial t} &= -g \frac{\partial h}{\partial x} +
fv, \\
\frac{\partial v}{\partial t} &= -g \frac{\partial h}{\partial y} -
fu, \\
\frac{\partial h}{\partial t} &= -h_{0} \left( \frac{\partial
u}{\partial x} + \frac{\partial v}{\partial y} \right).
\end{aligned}
\end{equation}
Taking $(Oy)$ perpendicular to the coast, and considering a solution
describing the above situation, {\it i.e.} $v=0$, $u=\xi(y) \exp
i(\omega t-kx)$, $h =\eta(y) \exp i(\omega t-kx)$, one obtains:
\begin{equation}
\begin{aligned}
u &=u_{0} \exp \left( -\frac{fy}{\sqrt{gh_{0}}} \right) \exp
i(\omega t -kx), \\ h &= u \sqrt{\frac{h_{0}}{g}}.
\end{aligned}
\end{equation}
The dispersion relation is given by ${\omega}^{2} = gh_{0}k^{2}$,
as for usual gravity waves. The characteristic length $L \equiv
\sqrt{gh_{0}} / f$ is the {\it Rossby radius 
of deformation}. At a mid-latitude $\lambda\sim
45^\circ$, 
one finds $L \sim 2200$ km for $h_{0} \sim 5$ km, while for a
shallow sea, {\em i.e.} $h_{0} \sim 100$ m, one
 rather has $L \sim 300$ km. 
The surface shape generated by the Kelvin waves is plotted in figure
\ref{surface}. 
One can notice that the surface undulation is
trapped in the vicinity of the coast, and its spatial extension 
in the direction of the ocean is typically of
order $L$. 

\begin{figure}
\psfig{file=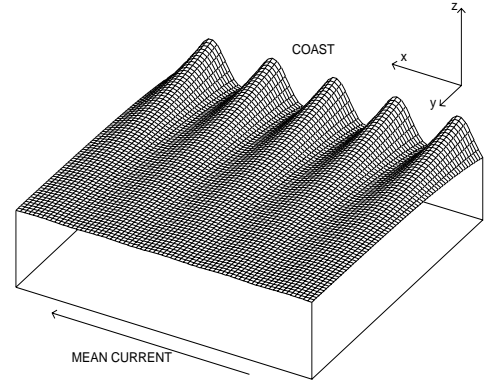,width=8.5cm,height=6.5cm}
\caption{ \label{surface} Surface shape generated by a Kelvin
wave.}
\end{figure}

The Kelvin waves are in fact easily observed, since 
the currents generated by tides are influenced by the Coriolis
force and give rise to them.
As a consequence, the coast is always
to the right of the flow direction (in the northern hemisphere).
On the oceanic basin scale, mean movements
are in this case an anticlockwise rotation around a point called {\it
amphidromic point}. 
This geometry is found in many places over the globe, the
rotation being clockwise in the southern hemisphere~\cite{openuniv2}.

\section{Conclusion}

Coriolis force effects become important as soon as the spatial
extension of the flow is important ($R_o \propto 1/L$).
This is the reason why the Earth's rotation considerably influences
the atmosphere and oceans dynamics.
We have presented in this paper several simple examples of
geophysical fluid dynamics. 
We hope it will be helpful for Mechanics teachers
to illustrate inertial effects with simple but physically relevant
examples.

\section*{Acknowledgments}
We thank P. C. W. Holdsworth for his kind help during
the preparation of the manuscript.

\end{document}

%% file: omega2.tex
\begingroup%
  \makeatletter%
  \newcommand{\GNUPLOTspecial}{%
    \@sanitize\catcode`\%=14\relax\special}%
  \setlength{\unitlength}{0.07bp}%
\scriptsize
\begin{picture}(3600,2160)(0,0)%
\special{psfile=omega2 llx=0 lly=0 urx=1028 ury=720 rwi=7200}
\put(3037,1847){\makebox(0,0)[r]{$\omega_-$}}%
\put(3037,1947){\makebox(0,0)[r]{$\omega_+$}}%
\put(2365,1532){\makebox(0,0)[l]{Tropical hurricanes}}%
\put(1978,447){\makebox(0,0)[l]{Low pressure centre ($G<0$)}}%
\put(389,447){\makebox(0,0)[l]{High pressure centre ($G<0$)}}%
\put(621,1239){\makebox(0,0)[l]{anticyclones}}%
\put(621,1356){\makebox(0,0)[l]{Extratropical}}%
\put(621,1532){\makebox(0,0)[l]{Extratropical cyclones}}%
\put(2365,1532){\makebox(0,0)[l]{Tropical hurricanes}}%
\put(1978,447){\makebox(0,0)[l]{Low pressure centre ($G<0$)}}%
\put(389,447){\makebox(0,0)[l]{High pressure centre ($G<0$)}}%
\put(621,1239){\makebox(0,0)[l]{anticyclones}}%
\put(621,1356){\makebox(0,0)[l]{Extratropical}}%
\put(621,1532){\makebox(0,0)[l]{Extratropical cyclones}}%
\put(1900,50){\makebox(0,0){$G/G_0$}}%
\put(100,1180){%
\special{ps: gsave currentpoint currentpoint translate
270 rotate neg exch neg exch translate}%
\makebox(0,0)[b]{\shortstack{$2 \omega/f  (f>0)$}}%
\special{ps: currentpoint grestore moveto}%
}%
\put(3450,200){\makebox(0,0){4}}%
\put(3063,200){\makebox(0,0){3}}%
\put(2675,200){\makebox(0,0){2}}%
\put(2288,200){\makebox(0,0){1}}%
\put(1900,200){\makebox(0,0){0}}%
\put(1513,200){\makebox(0,0){-1}}%
\put(1125,200){\makebox(0,0){-2}}%
\put(738,200){\makebox(0,0){-3}}%
\put(350,200){\makebox(0,0){-4}}%
\put(300,2060){\makebox(0,0)[r]{2}}%
\put(300,1767){\makebox(0,0)[r]{1}}%
\put(300,1473){\makebox(0,0)[r]{0}}%
\put(300,1180){\makebox(0,0)[r]{-1}}%
\put(300,887){\makebox(0,0)[r]{-2}}%
\put(300,593){\makebox(0,0)[r]{-3}}%
\put(300,300){\makebox(0,0)[r]{-4}}%
\end{picture}%
\endgroup
 

%% file: coriolis2.bbl
\begin{references}
\bibitem{note} For oceans, $R_e \sim 10^{11}$, with $
U \sim 1\ {\mathrm m}\cdot{\mathrm s}^{-1}$, $\nu\sim10^{-6}\
{\mathrm m}^2\cdot{\mathrm s}^{-1}$: the approximation is thus the
same.
\end{references}
